\documentclass[aps,prl,reprint]{revtex4-1}
\usepackage{graphicx}  
\usepackage{epstopdf}
\usepackage{bm}
\usepackage{amsmath} 
\usepackage{amssymb} 
\usepackage{color}
\usepackage{dcolumn}   
\begin{document}
\newcommand{\eps}[0]{\varepsilon}
\newcommand{\sxx}[0]{\sigma_{xx}}
\newcommand{\sxy}[0]{\sigma_{xy}}
\newcommand{\syy}[0]{\sigma_{yy}}
\newcommand{\FvK}[0]{F{\"o}ppl-von K{\`a}rm{\`a}n\,}

\title{Extreme Contractility and Torsional Compliance of Soft Ribbons under High Twist}
\author{Julien Chopin}
\email{julien.chopin@ufba.br}
\affiliation{Civil Engineering Department, COPPE, Universidade Federal do Rio de Janeiro, 21941-972, Rio de Janeiro RJ, Brazil}
\affiliation{Instituto de F{\'i}sica, Universidade Federal da Bahia, Campus Universit{\'a}rio de Ondina, rua Bar{\~a}o de Jeremoabo, BA 40210-340, Brazil}
\author{Romildo T. D. Filho}
\affiliation{Civil Engineering Department, COPPE, Universidade Federal do Rio de Janeiro, 21941-972, Rio de Janeiro RJ, Brazil}

\begin{abstract}
We investigate experimentally and model the mechanical response of a soft Hookean ribbon submitted to large twist $\eta$ and longitudinal tension $T$, under clamped boundary conditions. We derive a formula for the torque $M$ using the \FvK equations up to third order in twist, incorporating a twist-tension coupling.
In the stable helicoid regime, quantitative agreement with experimental data is obtained. When twisted above a critical twist $\eta_L(T)$, ribbons develop wrinkles and folds which modify qualitatively the mechanical behavior. We show a surprisingly large longitudinal contraction upon twist, reminiscent of a Poynting effect, and a much lower torsional stiffness. Far from threshold, we identify two regimes depending on the applied $T$. In a high-$T$ regime, we find that the torque scales as $\eta\cdot T$ and the contraction as $\eta^2$, in agreement with a far from threshold analysis where compression and bending stresses are neglected. In a low-$T$ regime, the contraction still scales as $\eta^2$ but the torque appears $T$-independent and linear with $\eta$. We argue that the large curvature of the folds now contribute significantly to the torque. This regime is discussed in the context of asymptotic isometry for very thin plates submitted to vanishing tension but large change of shape, as in crumpling.
\end{abstract}
\date{\today}
\maketitle
\section{Introduction}
Rods and Filaments are fundamental structures which play a key role in the mechanical behavior of many man made and biological materials, as well as large scale structures in civil and aeronautic engineering~\cite{bazant2002stability,wyart2008elasticity,Audoly2010,mackintosh1995elasticity}. At smaller scale, individual slender structures are widely used in MEMS~\cite{rogers2010materials}.
In the context of the locomotion at small Reynolds number, the beating of filaments appears as a common propulsion strategy shared by many small living organisms~\cite{purcell1977life,lauga2009hydrodynamics}, yielding non trivial fluid-structure dynamics~\cite{wiggins1998flexive,coq2008rotational,chopin2017dynamic}.

The understanding of the mechanics of slender solids has played a pivotal role for the development of the theory of elasticity, in particular, and the foundation of continuum mechanics, in general~\cite{timoshenko1953history}. One of the earliest result in elasticity is due to Coulomb's work on the torsion of circular rods. He established the celebrated formula for the torque $M$ which develops in a rod submitted to a twist rate $\tau$~\cite{Coulomb1784}:
\begin{equation}
    M = G J \tau,
    \label{eq:coulomb}
\end{equation}
where $G$ is the shear modulus, $J$ the twist moment which only depends on the geometry of the cross-section. Much later, Saint-Venant developed a linear elastic theory for the torsion of inextensible rods allowing to derive an exact formula for the twist moment for an arbitrary cross-section~\cite{SaintVenant1855}. 

In this paper, we investigate the nonlinear behavior of soft (i.e. extensible) elastic ribbon, characterized by a flat cross-section, when subjected to a large twist rate. Nonlinear mechanics of rods with circular cross-section has been the subject of numerous experimental and theoretical investigations~\cite{Audoly2010}. In most cases, the rod is assumed to be inextensible. Because of strong geometric nonlinearities, rods can exhibit various instabilities, even within the regime of Hookean (i.e. linear elastic) response. For example, an initially straight rod develops a localized helical buckling instability when submitted to a longitudinal tension and twisted above a critical torque~\cite{van2000helical}.

Richer set of instabilities have been investigated for inextensible ribbons. A ribbon is intermediate between a circular rod ($t \sim W \ll L$) and a thin plate $t \ll W \sim L $, where $t$, $W$, and $L$ are the thickness, the width, and the length, respectively. This hybrid geometry strongly impacts the mechanical behavior. As any filaments, ribbons are highly flexible and sustain large shape change in 3D for relatively small applied load. However, as a thin plate, they are subjected to a geometrical constraint which restrict the set of 3D shapes accessible at a low energy cost. The reason is given by Euler's Teorema Egregium : a flat surface forced into a shape with a non zero Gaussian curvature (i.e. with two non zero principal curvatures) necessary develops stretching. Since the energy cost for bending ($\sim h^3$) is much larger than for stretching ($\sim h$) as the thickness $h$ goes to zero, inextensible deformations, when possible, are largely favored. 

\begin{figure*}[ht]
\centering
\includegraphics[width = 14cm]{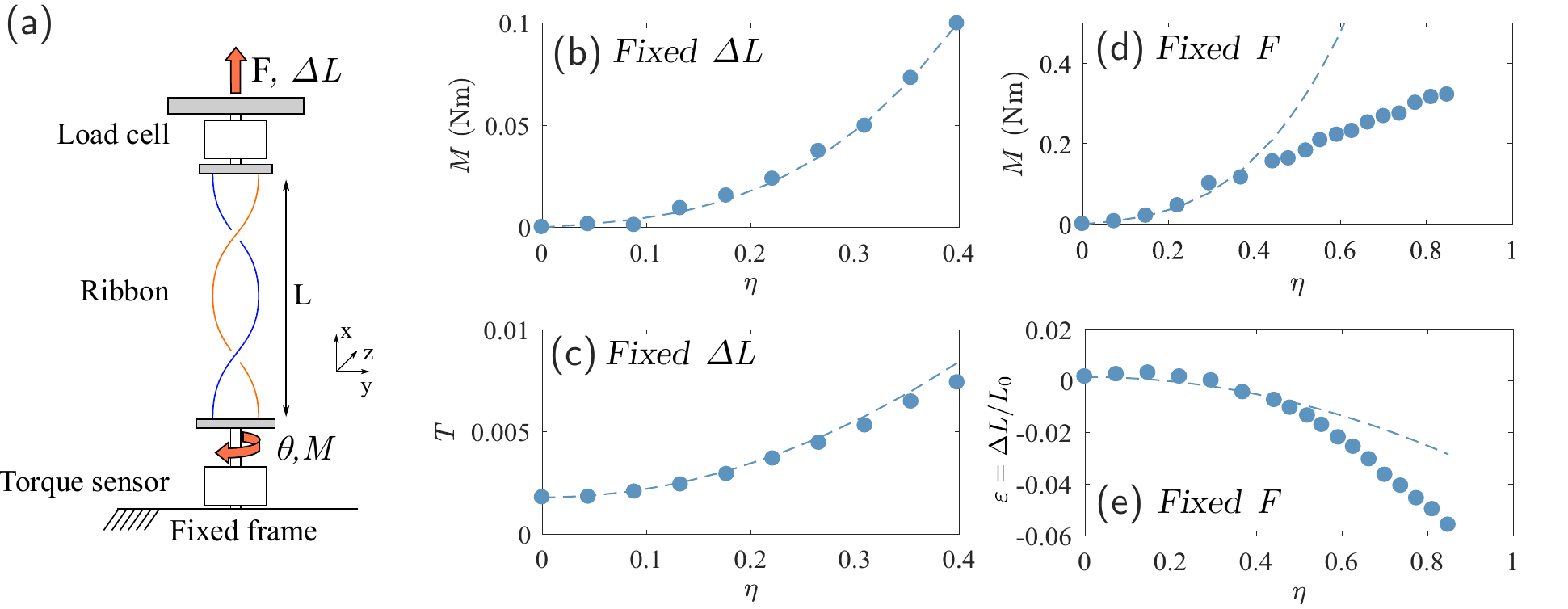}
\caption{(a) Schematics of the experimental setup. A ribbon is twisted by an angle $\theta$ under fixed load $F$ or fixed displacement $\Delta L$. The torque $M$ and normalized longitudinal tension $T = F/(EhW)$ are measured as a function of the normalized twist rate $\eta = \theta W/L$. Under fixed distance condition the nonlinear evolution of (b) $M$  and (c) $T$ are in excellent agreement with Eq.~\ref{eq:M} and  \ref{eq:T}, respectively. Under fixed load condition, (d) $M$ varies quasi linearly with $\eta$ at large twist. (e) A large contraction $\Delta L<0$ is observed and is nonlinear with $\eta$. Theoretical predictions given by Eq.~\ref{eq:M} and ~\ref{eq:eps} (dashed lines) do not capture the evolution of the torque (d) and the contraction (e) at large $\eta$.}
\label{Fig1}
\end{figure*}

Inextensible deformations are not always possible. A sheet under biaxial compression, as in the case of the packing of a sheet inside a small container, is an obvious example where stretching mode cannot be avoided. But, the sheet can crumple allowing to release a large fraction of the stretching energy~\cite{Blair2005,Witten2007}, the remaining part being  localized in a complex networks of folds. Thus, crumpled configurations can be interpreted as resulting from a trade-off to satisfy the boundary conditions and reducing the high stretching energy in favor of low bending energy. The folds are usually modeled as line-like, singular elastic structures~\cite{Lobkosvky1995,Fuentealba2015} which result from the interaction of more fundamental point-like singularities : the developable  cones, or d-cones~\cite{BenAmar1997,Boudaoud2000}.

The ribbons, considered as a thin plate, have been an interesting and fruitful playground to investigate the singular behavior of thin sheets under more or less complicated loading conditions~\cite{maurin2017solitary,yu2018bifurcations,morigaki2016stretching}. Most of the theoretical and numerical studies usually consider stiff materials under mild tension, thus, satisfying the condition of inextensibility~\cite{Sadowsky1930,Wunderlich1962,Dias2014,Dias2015}. Interestingly, curvature condensation resembling a crumpled state has been investigated in the context of the shape of Moebius bands~\cite{StarostinVanDerHeijden2007}. The folds network is ordered and is made of triangular facets whose vertices meet slightly outside the band. But, again, the shape remains entirely developable, thus, no regions are stretched which is markedly different from crumpling. 

When the inextensibility constraint is relaxed, it has been demonstrated that twisted ribbons under moderate tensile state exhibit a rich variety of morphologies that are controlled not only by the twist but also by the applied tension and the ribbon geometry ($t/W$ and $L/W$), yielding a 4D phase diagram~\cite{Chopin2013,Chopin2015,kudrolli2018tension}. This phase diagram, in the $\eta-T$ plane, exhibits a triple point at $T_c$ where three phase boundaries meet. At sufficiently small twist, the ribbon has a helicoid shape. For $T > T_c$, the helicoid is unstable against transverse wrinkling modes. The onset of the instability and the detailed morphology of the wrinkling structures are complex functions of the geometry and the loading~\cite{Chopin2013,Chopin2015,kudrolli2018tension}. Below $T_c$, ribbons destabilize above a $T$-dependent critical twist and develops longitudinal wrinkles, folds, loops, depending on the distance to the threshold. In a regime of large twist and moderate tension, the ribbon develops triangular facets connected by folds, reminiscent of the morphology found in Moebius bands~\cite{Korte2011}. However, it has been argued that, because of the small but finite tension, ribbons develop a so-called e-helicoid shape, which consists of the interaction of excess cones, or e-cones~\cite{chopin2016disclinations}. E-cones are point-like, singular, elastic structures but, unlike d-cones, they do not preserve the metric~\cite{Muller2008,Efrati2015,seffen2016fundamental}.

In this paper, we explore the nonlinear mechanics of soft ribbons subjected to a moderate longitudinal load and small to large twisting rate. We essentially focus on the longitudinal strain and torque response, thus providing new insights compared to the more morphology-oriented investigations carried out so far. 
We show experimentally that ribbon exhibits a strongly nonlinear torque and strain response, even for moderate twist rate. We develop a nonlinear torsion model of soft ribbon based on the \FvK equations which quantitatively captures the mechanical response, up to a critical threshold above which a longitudinal wrinkling instability develops. In the last part of the paper, we analyze and model the mechanical response at high twist in regimes where the wrinkling instability plays a major role. Far from threshold, we identify two regimes depending on the applied $T$. It is then argued that the transition between the two occurs when the curvature of the wrinkles and folds is large enough to contribute significantly to the elastic energy. Finally, these regimes are modeled and interpreted in the light of recent theoretical approaches addressing the morphology and mechanics of highly wrinkled thin plates far from threshold.

\section{Experiments}

The schematics of the experimental setup is shown in Figure~\ref{Fig1}(a). An initially flat ribbon is twisted by an angle $\theta$ and stretched longitudinally by a force $F$, using clamped boundary conditions. The ribbon is composed of cellulose acetate with Young's modulus $E\simeq 2.2$\,GPa and Poisson ratio $\nu = 0.35\pm 0.05$. The thickness is  $t=256\,\mu$m, width $W=35\,$mm and length $L_0 = 331\,$mm. The force $F$ is measured by mean of a load cell mounted on the upper clamp whose vertical position can be adjusted to vary the distance $L$ to the bottom clamp. The clamp displacement is defined as  $\Delta L = L - L_0 $ and is measured by mean of a micrometer. When $\Delta L = 0$, the ribbon is in its reference configuration, thus not stretched nor compressed. The torque $M$ is measured by a torque sensor placed at the lower clamp which can rotate around the longitudinal axis but cannot translate in the vertical direction. Two loading conditions are considered : 1- A fixed load condition where $F$ is kept fixed while $\Delta L(\theta)$ is a free parameter, 2- A fixed displacement condition where $\Delta L > 0$ is fixed and $F(\theta)$ is a free parameter. 

We now introduce nondimensional quantities. The normalized twist rate is defined as : 
\begin{eqnarray}
\eta = \tau W = \theta \frac{W}{L_0}
\end{eqnarray}
In the linear elastic regime where Eq.~\ref{eq:coulomb} applies ($\eta \ll 1$), $\eta$ is a measure of the characteristic strain developing in the rod upon twist. Further, the longitudinal strain $\eps$ and the normalized longitudinal tension $T$ are defined by 
\begin{eqnarray}
\eps &=& \frac{\Delta L}{L_0},\label{eq:eps_def}\\
T &=& \frac{F}{E h W}, \label{eq:T_def}
\end{eqnarray}
respectively.

In Fig.~\ref{Fig1}(b) and \ref{Fig1}(c), we show the evolution of the torque $M$ and tension $T$ with $\eta$ over a large range of twist. The displacement $\Delta L$, or, equivalently $\eps$, is kept fixed at $\eps = 1.8\times 10^{-3}$. The shape of the ribbon is a stretched helicoid which remains stable over the range of twist applied. On the application of an initial strain, the ribbon deforms longitudinally and develops a longitudinal tension $T = \eps$ at $\eta = 0$ as seen in Fig.~\ref{Fig1}(c). No offset is observed in the torque, as one would intuitively guessed. Then, upon increasing the twist, we find that the tension and torque increase non linearly with $\eta$, even for relatively small twist $\eta \approx 0.1$. The linear response expected at small $\eta$ could not been observed because it is below the accuracy of our measurements.

The nonlinear increase of the torque can be seen as an effective shear modulus $G_{eff} = M/(J\,\tau)$ increasing with the twist. This behavior is analogous to a strain hardening effect commonly observed in  crosslinked polymers like rubber submitted to a large shear~\cite{treloar1975physics}. The strain hardening is a nonlinear material property whose microscopic origin is usually subtle and relies a variety of processes depending on the material considered. However, the ribbon used in our experiment are loaded within their linear elastic response and no significant plastic deformation occur when $\eta < 0.4$. Therefore, the strain hardening observed here can hardly be attributed to complex material nonlinearities.  In parallel, the ribbon develops a nonlinear tension which is coupled with the twist. Such coupling has already been investigated but for very soft materials submitted to finite deformation~\cite{ghatak2005solenoids}. Later on, we explain these behaviors as resulting from geometrical nonlinearities arising at high twist.

Next, we investigate the effect of the loading conditions. The ribbon is now twisted under fixed load conditions with $T = 1.6\times 10^{-3}$. Note that, initially, the ribbon is stretched to almost the same extent as previously, under fixed distance condition. In Fig.~\ref{Fig1}(d) and ~\ref{Fig1}(e), we plot the torque $M$ and the tension $T$, respectively. Initially, the $M(\eta)$ curve increases nonlinearly in a similar fashion as under a fixed distance condition. But, at larger twist, we observe a change of trend characterized by an almost linear increase. This change of behavior coincides with a loss of stability of the helicoid shape and signals the effect of instability on the mechanical behavior. Incidentally, we observed an effective strain softening corresponding to a lower torsional stiffness or, equivalently, a lower effective shear modulus, compared to the projected nonlinear behavior without instability (blue dashed line). Similarly, in Fig.~\ref{Fig1}(e), we observe an enhanced contractility of the ribbon ($\Delta L < 0)$ at large $\eta$.

Interestingly, the large contraction is reminiscent of the Poynting effect~\cite{poynting1909pressure} which corresponds to the tendency of sheared materials to elongate or contract normal to the shear plane.  While a quantitative explanation of this effect is still under debate, it is generally accepted that it manifests itself in regime of large deformation~\cite{janmey2007negative,mihai2011positive,baumgarten2018normal}. It is however remarkable that a large Poynting effect can be observed in linear elastic structures under small deformation but finite rotation.  

\section{Saint-Venant linear torsion theory }
We now discuss the results of the Saint-Venant torsion theory to highlight the specificity of the mechanical behavior observed in extensible ribbons. In Saint-Venant theory, materials have a Hookean elasticity. Consistently, all the strains are small, thus, $T \ll 1$ and $\eta \ll 1$. The angle of torsion is also small $\theta \ll 1$ which implies that all the nonlinear terms in the definition of the strain are neglected~\cite{LandauLifshitz1986}. Finally, the rod is assumed inextensible and infinite.

We denote $u(x,y,z)$, $v(x,y,z)$, and $w(x,y,z)$ the components of the displacement in the $x$, $y$, and $z$ directions, respectively. When a twist rate $\tau$ is applied to the rod, the displacement field satisfying the equation of mechanical equilibrium  has the following form~\cite{SaintVenant1855,love2013treatise}:
\begin{eqnarray}
    u(x,y,z) &=& \tau \phi(y,z), \label{eq:u_StV}\\
    v(x,y,z) &=& - \tau x z, \\
    w(x,y,z) &=& \tau x y,
\end{eqnarray}
where the warping function $\phi(y,z)$ is a harmonic function satisfying boundary conditions consistent with a stress free condition at the rod lateral surface. For a circular (isotropic) cross-section of radius $R = W/2$, the cross-section remains flat under torsion ($\phi = 0$) and the twist moment $J_{iso} = G W^4 \pi/32$. For a rectangular (anisotropic) cross-section, the longitudinal displacement is not zero ($\phi \neq 0$) and the cross-section warps. There is no simple analytical expression for $\phi$ and $J$, but in the limit $W\gg t$, the twist moment is given by~\cite{timoshenko1951theory}:
\begin{equation}
    J_{ani} = \frac{1}{3}t^3W = \frac{1}{3}\left(\frac{t}{W}\right)^2 (h\,W)^2.
    \label{eq:Jani}
\end{equation}
A ribbon is much more compliant that a rod as its moment is smaller by a factor $(t/W)^2$ than a circular rod, for the same cross-section area. From the parameters used in the experiment, twisting a ribbon by $\eta = 0.2$ generates a typical torque $M \approx 10^{-3}$\,N.m, according to Eq.~\ref{eq:Jani}. From Fig.~\ref{Fig1}(b) and \ref{Fig1}(c), we clearly see that the Saint-Venant theory predicts a torque at least one order of magnitude smaller than what is measured.

Further, the change of ribbon length cannot be addressed within the original Saint-Venant theory because of the inextensible condition. Even if a degree of extensibility is incorporated, adding a term $\varepsilon\,x$ in Eq.~\ref{eq:u_StV}, this would result in a positive strain $\eps = T>0$. This additional degree of freedom in the kinematics would capture an initial positive offset of the $\eps(\eta)$ curve but not the nonlinear trend and the change of sign at large $\eta$. In the experiment, we observe a large contraction which is an order of magnitude larger than the applied tension. For example, when twisted by 1.5 turn, corresponding to $\eta \approx 0.9$, the ribbon does not break or yield but contracts by almost 10\%. It is interesting to see that this value obtained by a simple elastic ribbon is of the order of the performance measured in artificial muscles~\cite{haines2014artificial}.

\begin{figure*}[ht]
\centering
\includegraphics[width = 14cm]{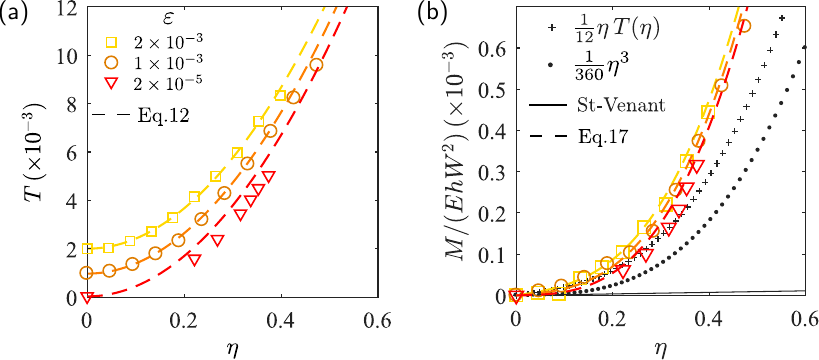}
\caption{Nonlinear evolution of (a) the tension $T$ and (b) the torque $M$ with twist $\eta$ under fixed distance condition for different imposed pre-strains $\eps = 2\times10^{-5}$, $10^{-3}$ and $2\times10^{-3}$. Quantitative agreement with theoretical predictions of the nonlinear torsion theory (dashed line) given by Eq.~\ref{eq:T} and Eq.~\ref{eq:M} is obtained without adjusting parameters. For comparison, the linear Saint-Venant model (solid lines) is shown in (b) along with nonlinear contributions to the torque ($\bullet$, $+$).}
\label{Fig2}
\end{figure*}

\section{Nonlinear torsion model of soft ribbons}

We will now develop a nonlinear model for the torsion of ribbons using the \FvK (FvK) which describes the equilibrium configurations of thin elastic plates under large deflections. The FvK equations are a set of nonlinear, partial differential equations where the unknown fields are the deflection of the mid-surface $w(x,y)$ in the $z$ direction and the planar stress $\sxx(x,y)$, $\syy(x,y)$, and $\sxy(x,y)$ evaluated at the mid-plane surface ($z=0$)~\cite{Audoly2010}. The FvK equations can be derived formally as an expansion of the equations of equilibrium of 3D elasticity in power of the ratio of the thickness over the width of the plate.  This reduction of dimensionality relies on a series of approximations. Chief among them are : 1-the small slope approximation implying that configurations do not depart excessively from a plane; 2- some nonlinear terms in the Green-Lagrange strain tensor are neglected
; 3- the cross-section remains normal to the mid-plane upon loading which is also known as the Kirchhoff hypothesis. 

Using the Kirchhoff hypothesis, we can obtain the in-plane displacement at a distance $z$ from the mid-plane. At linear order in $z$, $u(x,y,z)= u_1(x,y) - z \partial w/\partial x$ and $v(x,y,z)= v_1(x,y) - z \partial w/\partial y$, where $u_1$ and $v_1$ are the in-plane displacement along the $x$ and $y$ axis, respectively; evaluated at $z=0$~\cite{Audoly2010}. We can now give the general form of the displacement fields for a ribbon in the context of the FvK equations. For a helicoid with a twisting rate $\tau$, the deflection in the small slope limit reads :
\begin{eqnarray}
w(x,y)&=& \tau x y.\label{eq:w}
\end{eqnarray}
Thus, at linear order in $z$, we have:
\begin{eqnarray}
u(x,y,z)&=& u_1(x,y) - \tau z y,\label{eq:u}\\
v(x,y,z)&=& v_1(x,y) - \tau z x,\label{eq:v}.
\end{eqnarray}
Comparing with the kinematics in the Saint-Venant theory, we find that $\phi(y,z) =  - z\,y$. The longitudinal and transverse displacements, $u_1$ and $v_1$, respectively, are two additional terms which account for the in-plane stretching of the mid-surface. To linear order in $\eta$, we expect that $u_1 = \eps\,x$ and $v_1 = -\nu \eps \, y$. This would correspond to a longitudinal displacement in response to a load $T$ and the corresponding contraction in the transverse direction by a Poisson effect. With the FvK equations, geometrical nonlinearities are now present and leads to new terms in the stress and the displacement fields. In the following, we will see that these new terms are not just small corrections to the linear solution.

Solving the FvK equations assuming the helicoid geometry, we obtain the stress components \cite{Green1937,Coman2008,Chopin2013,Chopin2015} : 
\begin{equation}
 \sigma_{yy}= \sigma_{xy}=0, \quad \sxx/E = \eps + \frac{1}{2}\eta^2(y/W)^2.
\label{eq:sss}
\end{equation}
Thus, for a stretched helicoid, in the small slope limit, there is no shear and transverse stresses, at the leading order in the ribbon thickness ($h/W = 0$). At this order, the longitudinal stress is non-zero, $x$-independent and is a parabolic function of $y$. Using Eqs.~\ref{eq:u} and ~\ref{eq:v}, shear stress of order $\mathcal{O}(h^2/W^2)$ will arise and will contribute to the bending energy of the plate, as we will see later on. The longitudinal strain $\eps$ depends in general of $\eta$ and $T$. For a fixed load condition, the corresponding strain is :
\begin{equation}
\eps(\eta,T) = T- \frac{\eta^2}{24}.
\label{eq:eps}
\end{equation}
and, for a fixed displacement condition, the corresponding tension is :
\begin{equation}
T(\eta,\eps) = \eps + \frac{\eta^2}{24}.
\label{eq:T}
\end{equation}
The in-plane displacements at $z=0$ are given by :
\begin{eqnarray}
 u_1(x,y) &=& \eps\,x + \frac{1}{2}\,\eta^2 (y/W)^2\,x \label{eq:u1}\\
 v_1(x,y) &=& -\nu \eps\,y - \frac{\nu}{6}\, \eta^2\,(y/W)^3\,W.\label{eq:v1}
\end{eqnarray}
In Eqs.~\ref{eq:u1} and \ref{eq:v1}, we recover the Poisson effect at linear order in $\eps$ and nonlinear effects ($\sim \eta^2$) which arises from the nonlinear terms in the strain tensor. 

To obtain the torque, we use an energetic approach defining $M$ as : 
\begin{equation}
M = \frac{\partial U}{\partial \theta} \bigg|_{T},
\label{eq:M_def}
\end{equation}
where $U/(EhWL) = U_s + U_b -\eps\,T$ is the total energy of the ribbon including the work of the load, the stretching energy $U_s$ and the bending energy $U_b$. The stretching energy is simply given by $U_s/(EhWL) = 1/(2E) \int  \sigma_{xx}^2\,dy/W$ since the other components of the stress are zero. For a helicoid, the mean curvature is zero and the Gaussian curvature is $K \approx -\tau^2$. Thus, the  bending energy can be easily calculated $U_b/(EhWL)  =-(B/h)\,K/2$, where $B = Gh^3/[6(1-\nu)]$ is the bending modulus ~\cite{LandauLifshitz1986}. After integration along the transverse direction, the total energy  reads :
\begin{equation}
\frac{U}{EhWL}= \frac{1}{1440} \eta^4 + \frac{1}{24} T\eta^2 -\frac{1}{2}T^2+ \frac{1}{6}\frac{G}{E}\left(\frac{h}{W}\right)^2 \eta^2,
\label{eq:FvKEnergyHel}
\end{equation}
Using Eq.~\ref{eq:M_def} and Eq.~\ref{eq:FvKEnergyHel}, we obtain the expression of the torque :
\begin{equation}
\frac{M(\eta,T)}{EhW^2} = \frac{1}{3}\frac{G}{E}\left(\frac{h}{W}\right)^2 \,\eta + \frac{1}{360} \eta^3   + \frac{1}{12}\,\eta T.
\label{eq:M}
\end{equation}
The torsional stiffness defined as $C = M /\tau$ is given under fixed load condition by :
\begin{equation}
\frac{C(\eta,T)}{EhW^3} = \frac{1}{3}\frac{G}{E}\left(\frac{h}{W}\right)^2  + \frac{1}{360} \eta^2   + \frac{1}{12} T,
\label{eq:C}
\end{equation}
The corresponding formula for the torque and torsional stiffness for a fixed displacement condition can be expressed in terms of $\eps$ and $\eta$ by substituting in Eq.~\ref{eq:M} and ~\ref{eq:C} the expression of $T$ given by Eq.~\ref{eq:T}. 

\begin{figure*}
    \centering
    \includegraphics[width = 10.5cm]{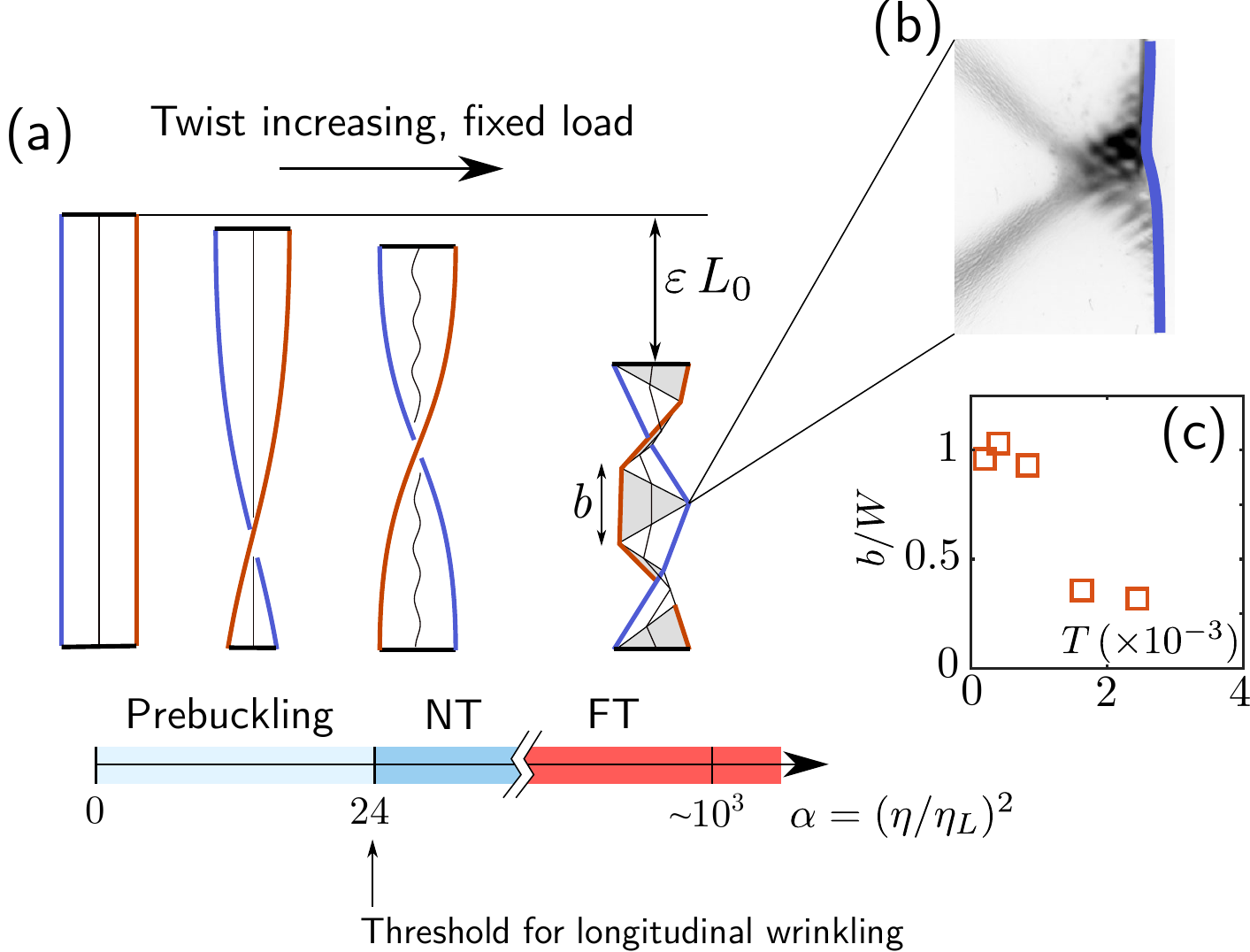}
    \caption{Schematics of the morphologies observed when twisting a ribbon under constant load as a function of $\alpha =24 (\eta/\eta_L)^2$. The helicoid shape is stable below threshold for the longitudinal wrinkling instability ($\eta<\eta_L$, or $\alpha <24$. For $\alpha > 24$, the helicoid is wrinkled near threshold (NT), and far from threshold (FT), the ribbon shows triangular facets. (b) Large deformation (black marks) are observed at the folds between two facets. (c) Evolution of the facets aspect ratio $b/W$ with the tension.}
    \label{fig:origami}
\end{figure*}
Let us now discuss the three terms in Eq.~\ref{eq:C}. The first term is the linear Saint-Venant contribution. By comparison with the two other terms, we find that the Saint-Venant model is an accurate model for $T \ll (h/W)^2$ and $\eta \ll h/W$. These limits are easily reached for a ribbon with typical aspect ratio $h/W \sim 10^{-2}$ yielding $T \sim 10^{-2}$ and $\eta \sim 10^{-2}$. This explains why the linear regime is not easily observed in soft ribbon. The second term is a nonlinear contribution first obtained by Green in the limit of small but finite $\eta$ and $T \ll \eta^2$ \cite{Green1937}. The second nonlinear contribution proportional to $T$ has been first considered by Buckley~\cite{buckley1914} and derived rigorously by Biot~\cite{biot1939increase} in the limit of small but finite $T$ and $\eta \ll \sqrt{T}$, a regime complementary to that considered by Green.  Therefore, our derivation captures three regimes in a unique formula. It is noteworthy that both nonlinear contributions, previously considered independently, play a significant role to the overall torsional response of soft ribbon at large $\eta$.


To validate further our model, we measure the torque and contraction varying $\eps$ under fixed distance condition. In Fig.~\ref{Fig2}(a) and (b), we plot the evolution of the tension and the torque with the twist for three different fixed strains $\eps = 0.02\times10^{-3},\,1.0\times10^{-3},\,2.0\times10^{-3}$. The combination of strain and twist is such that the ribbon remains stable against a transverse wrinkling instability~\cite{kudrolli2018tension}. In Fig.~\ref{Fig2}(a), the tension show a vertical shift and quadratic evolution with $\eta$. Excellent agreement in obtained with Eq.~\ref{eq:eps} (dashed lines) without any fitting parameters. In Fig.\ref{Fig2}(b), we plot the corresponding data for the torque. Eq~\ref{eq:M} (dashed line) quantitatively captures the nonlinear increase with the twist and the shift due the coupling with the tension. The prediction of the Saint-Venant model (solid line) fails to capture not only the nonlinearities and the tension-induced shift but also the overall magnitude of the torque.

\section{Twisted ribbon, ordered crumpling, and self-organized origami}

As already shown in Fig.~\ref{Fig1}(d) and (e), large deviations from the predictions are observed in the experimental data for large twist. These deviations occur because the ribbon develops a longitudinal wrinkling instability which strongly affect the shape and the stress distribution.

The series of shape changes observed in ribbons due to the longitudinal wrinkling instability are schematically represented in Fig.~\ref{fig:origami}(a). This mechanical instability has been studied in detail in previous works~\cite{Chopin2013,Chopin2015,dinh2016cylindrical}. At small twist, the base state of the ribbon is a stretched helicoid whose shape is given by Eq.~\ref{eq:w} and stress field by Eq.~\ref{eq:sss}.  Below a critical tension $T_c$ which scales linearly with $h/W$, a longitudinal wrinkling instability develops for $\eta > \eta_L$. The instability is driven by a longitudinal compression in the central part of the ribbon. A linear perturbation analysis allows to capture the threshold~\cite{Coman2008,Chopin2013}:
\begin{equation}
    \eta_L = \sqrt{24\,T}+c(\nu) (h/W),
    \label{eq:etaL}
\end{equation}
where $c(\nu) \sim 10$ is a nondimensional prefactor weakly depending on the Poisson ratio. 
It is convenient to introduce a confinement parameter $\alpha$ defined as :
\begin{equation}
\alpha \equiv \frac{\eta^2}{T} \approx 24\, \left(\frac{\eta}{\eta_L}\right)^2.
\label{eq:alpha}
\end{equation}
The confinement parameter can be interpreted as the ratio of the geometrical strain over the mechanical strain. It is also a measure of the distance from the threshold of longitudinal instability. 

As the twist is increased above threshold ($\alpha > 24$), the width $d$ of the wrinkling zone expands laterally towards the ribbon long edges.  Far from threshold ($\alpha \gg 24$), the wrinkling pattern exhibits a symmetry breaking along with a gradual localization of the elastic energy leading to the formation of a triangularly faceted helicoid~\cite{chopin2016disclinations}. 
It has been argued that the singularities observed in an extensible ribbon under twist are e-cone because they develop in a stretched region near the edge. The resulting structure is thus called an e-helicoid as opposed to a crumpled sheet where d-cones interact. 
As shown in the picture in Fig.~\ref{fig:origami}, we observe near the edge localized plastic deformations at the intersection of two folds while the rest of the ribbon undergoes reversible deformations. These marks demonstrate the development of large strain in localized regions.

The e-helicoid is an interesting structure which forms spontaneously as if one crumples an elastic sheet but the resulting network of folds is highly ordered like in origami. The morphology of this intermediate structure between origami and crumpled sheet have been studied experimentally and theoretically but their full understanding remains elusive~\cite{Chopin2013,dinh2016cylindrical}. To the best of our knowledge, the effect of the wrinkling instability and the formation of facets on the torsional and contractile response of ribbons have not been addressed experimentally and no explicit predictions for the corresponding torque stiffness and contractile response are available.

\begin{figure*}
    \centering
    \includegraphics[width = 14cm]{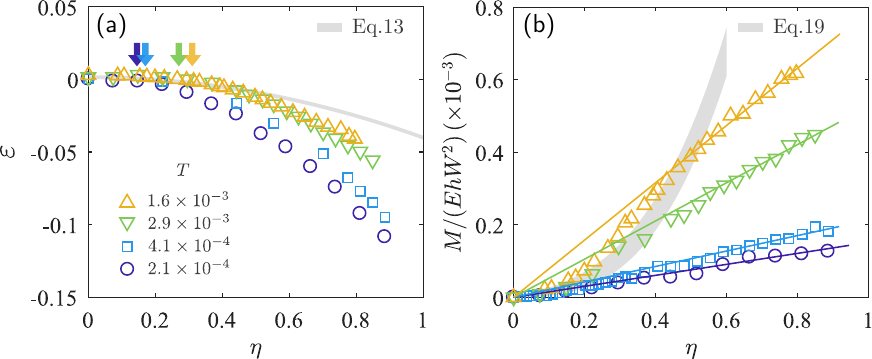}
    \caption{Nonlinear evolution of (a) the strain $\varepsilon$ and (b) the torque $M$ with twist $\eta$ under fixed load condition for different imposed tension. 
    Quantitative agreement with theoretical predictions of the nonlinear torsion model (gray area) given by Eq.~\ref{eq:eps} and Eq.~\ref{eq:M} only holds for $\eta < \eta_L(T)$ where the helicoid is stable. Predictions underestimate the contraction and overestimate the torque when $\eta>\eta_L$, especially at small $T$. The torque exhibits a linear dependence with $\eta > \eta_L$ (solid lines) with a $T$-dependent slope.}
    \label{Fig3}
\end{figure*}

\section{Ribbon torsional stiffness and contractility far from threshold}
We turn to the torsional stiffness and contractility in large twist regimes ($\alpha \gg 1$) and investigate the effect of the wrinkling instability. In Fig.~\ref{Fig3}(a), we plot the $\eps(\eta)$ curves varying the tension $T$ in the range $2\times 10^{-4}$ to $3 \times 10^{-3}$. We find that the contraction (measured positively) increases nonlinearly with $\eta$ and that all the curves lie systematically below the predictions given by Eq.~\ref{eq:eps}. The onset of deviations from Eq.~\ref{eq:eps} is found to be $T$-dependent which is fairly well captured by the longitudinal instability $\eta_L$ (see arrows). Unlike the pre-wrinkling and near threshold (NT) regime, the strong dependence of the contraction with the tension is a new qualitative feature observed far from threshold.

In Fig.~\ref{Fig3}(b), we plot the corresponding $M(\eta)$ for the same applied tension as in Fig.~\ref{Fig3}(a). In the far from threshold regime, we find the torques curves lie far below from the prediction for a stretched helicoid (grey area). The torque response is strongly depending on the tension unlike what is predicted by Eq.~\ref{eq:M}. The deviations are the largest for the smallest tension. Interestingly, a linear relation $M \sim \eta$ is observed in the post-buckling regime with a slope decreasing with the tension. We emphasize that this linear response is not captured by Eq.~\ref{eq:M} where a linear term with $\eta$ and $T$-dependent slope is present ($\sim T \eta/12$). This term is subdominant at large twist compare to the cubic term ($\sim \eta^3$). Indeed, for $\eta = 0.8$ and $T = 1.6\times 10^{-3}$, we have $\eta T/12 \approx 0.15 \times 10^{-3}$ which is much smaller that the measured torque $M/(EhW^2) \approx 0.6 \times 10^{-3}$.

\paragraph{High-T regime.} To model the torsional response and contractility of the ribbon in the wrinkling regime ($\eta>\eta_L$), we use a recent far from threshold (FT) approach which has been proven particularly useful to predict the wrinkling structure of ultra-thin sheet~\cite{Davidovitch2011}. While a standard linear stability analysis is valid for configuration sufficiently close to threshold allowing to take the pre-buckling state as a reference configuration~\cite{Chopin2013}, a FT approach assumes that, in the limit of vanishing thickness, the wrinkles completely relax the compression at no significant bending cost, thus strongly affecting the stress field. Unlike previous tension field theory~\cite{Mansfield2005}, this approach gives prediction for the extension of the unstable region and the morphological of the wrinkles far from threshold. To allow a quantitative analytical approach in the case of twisted ribbon, an ansatz for the stress field far from threshold has been proposed ~\cite{Chopin2015}: 
\begin{equation}
\sigma^{FT}_{xx}/E = \begin{cases} \frac{1}{2}\, \eta^2\, (y^2-d^2)/W^2, & \mbox{if } |y|>d \\ 0, & \mbox{if } |y|<d \end{cases}
\label{eq:sigmaFT}
\end{equation}
Outside the central wrinkled zone $|y|>d$ where the stress is zero, the stress field consists of longitudinal tensile component distributed as a parabola along the $y$ axis, like in the pre-buckled state, thus satisfying the equation of in-plane equilibrium. The wrinkled zone extension is set by the condition of vertical equilibrium $\langle \sigma^{FT}_{xx} \rangle_y /E = T$ where $ \langle . \rangle_y $ represents the average along the $y$ axis. Using Eq.~\ref{eq:sigmaFT}, we thus obtain an implicit relation between $d$, $\eta$, and $T$~\cite{Chopin2015}:
\begin{equation}
1-12\,\left(\frac{d(\eta,T)}{W}\right)^2+16\,\left(\frac{d(\eta,T)}{W}\right)^3= \frac{24}{\alpha(\eta,T)},
\label{eq:mecaeq}
\end{equation}
 When $\alpha \gg 1$, it can be easily shown that $d$ tends to $W/2$ meaning that the wrinkling structure invades the entire width of the ribbon. Because  the stress in the wrinkled region is zero, narrow bands under tensile stress develop near the edges and bear all the load. 
In this limit, the contraction is now given by~\cite{Chopin2015}:
\begin{equation}
\eps \approx -\frac{\eta^2}{2}\, \left(\frac{d(\eta,T)}{W}\right)^2
\end{equation}
To obtain the torque $M$ for $\alpha \gg 1$, we calculate the energy of the ribbon far from threshold. The work of the load $\sim T^2 \alpha$ is dominant compared to the stretching energy $T^2\,\sqrt{\alpha}$. Thus, in the FT limit ($\alpha \gg 1$ and $h \ll W$), the total energy of the ribbon reduces to : 
\begin{equation}
U_{FT}/(EhWL) \approx \frac{1}{8} \eta^2 T
\label{eq:U_FT}
\end{equation}
We obtain for the longitudinal strain: 
\begin{equation}
\eps_{FT} = - \frac{1}{8} \eta^2,
\label{eq:eps_FT}
\end{equation}
and for the torque and torsional stiffness
\begin{eqnarray}
 M_{FT}/(EhW^2) = \frac{1}{4} \eta T,\label{eq:M_FT}\\ 
 C_{FT}/(EhW^3) = \frac{1}{4} T,\label{eq:C_FT}
\end{eqnarray}
where we used the fact that $d$ tends to $W/2$ far from threshold.
\begin{figure}
\centering
\includegraphics[width =8cm]{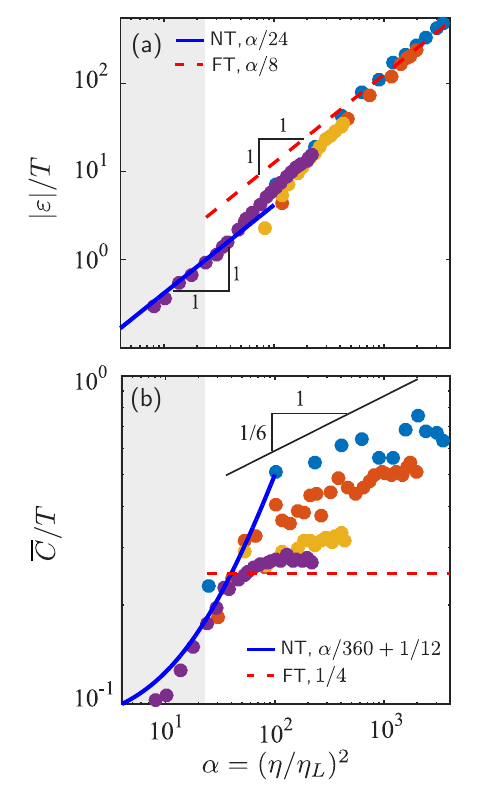}
\caption{(a) Evolution of  $|\varepsilon|/T$ with $\alpha$ for tensions $T = 2.1 \times10^{-4}$, $4.1 \times10^{-4}$, $1.6 \times10^{-3}$, and $2.3\times10^{-3}$. We observe a transition from the scaling $\alpha /24$ (blue line) to $\alpha /8$ at $\alpha \approx 10^2$ consistent with theoretical predictions given by Eqs.~\ref{eq:eps} and \ref{eq:eps_FT}. (b) Evolution of the ratio of the normalized torsional stiffness $\overline{C} = C / (EhW^3)$ over the tension $T$ with the confinement parameter $\alpha$ for the same tensions as in (a). In the post-buckling regime $\alpha > 24$, we find that $\overline{C}/T$ is mostly $T$-dependent.}
\label{Fig4}
\end{figure}

In the light of the predicted scalings, we now plot in Fig.~\ref{Fig4}(a) the contraction normalized by the tension as a function of the confinement parameter $\alpha$. The grey area ($\alpha < 24$) corresponds to the regime where the ribbon has a stable helicoid shape. In this regime the evolution of the contraction is well captured by the near threshold (NT) analysis (solid blue line) predicting that $|\eps|/T = \alpha /24$ (see Eq.~\ref{eq:eps}). Then, we identify a transition from NT to FT for $\alpha \sim 10^2$. This transition appears to be rather $T$ independent. For $\alpha > 10^2$, we find that the FT analysis provides a quantitative prediction for the contraction.

\begin{figure}
\centering
\includegraphics[width = 6cm]{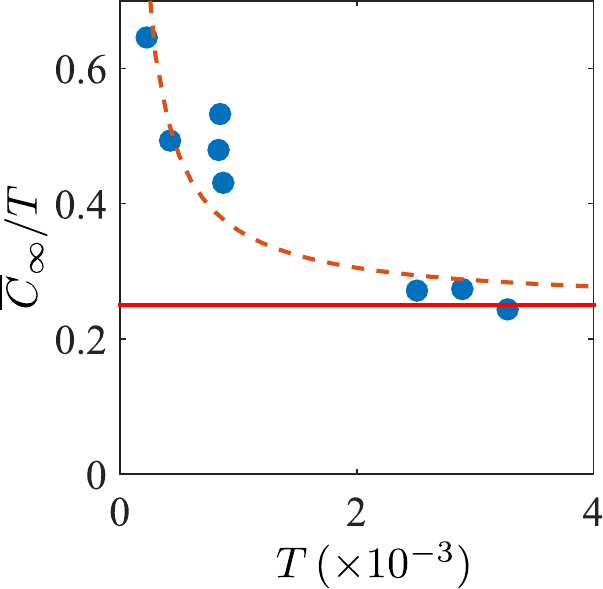}
\caption{Evolution of the asymptotic value of $\overline{C}_{\infty}/T$ for $\alpha\gg 24$. At large $T$, $\overline{C}_{\infty}/T$ tends to the value $1/4$, predicted by the FT approach.}
\label{Fig7}
\end{figure}

The corresponding measurements of the torque as a function of $\alpha$ are shown in Fig.~\ref{Fig4}(b). Here, we plot the normalized stiffness $\overline{C} = C / (Ehw^3)$ over the tension. In the regime of stable helicoid (grey area), the $\overline{C}(\alpha)/T$ curve is well captured by the NT analysis (solid blue line) which predicts $\overline{C}/T = \alpha /360 + 1/12$ (see Eq.~\ref{eq:C}). In the wrinkling regime, the torsional response appears to be more complex than the contraction response. For $\alpha > 10^2$, the torque deviates from the NT prediction but the trend is observed to depend on the applied tension unlike the contraction response. More precisely, the torque is found to globally decrease with $T$ and exhibits a weak dependence with $\alpha$ which can be captured empirically by a power law $C \sim \alpha^{p}$ with a small exponent $p$. The scaling $C \sim \alpha^{1/6}$ is shown as a guide for the eyes. In a first approximation, this dependence is however neglected, thus,  we assume that a plateau $C_{\infty}/T$ is reached at large $\alpha$. As shown in Fig.~\ref{Fig7}, the plateau decreases with $T$ and reaches asymptotically the value $1/4$. Therefore, we find that the FT analysis correctly predict the torque (see Eq.~\ref{eq:C_FT}) for sufficiently large tension.

This behavior can be interpreted as follows. In this high-$T$ regime, stretching energy most likely dominates bending energy, thus, our FT approach where the bending contribution of the folds is neglected in the stress field provides an accurate torsion model. However, at smaller tension, the stress field may be qualitatively different from the proposed form given by Eq.~\ref{eq:sigmaFT} as the large curvatures along ridges and vertices now contribute significantly to the stress. The change of regime from high to low tension is also observed  with the evolution of the triangular pattern, as shown Fig.~\ref{fig:origami}(c). When decreasing the tension, we found a transition at $T > 1 \times 10^{-2}$ where the aspect ratio sharply increases until reaching a plateau $b/W \approx 2$. Note that we do not observe a significant influence of the twist on the triangle shape. No higher value of $b/W$ could be obtained by reducing further the tension.

\paragraph{Low-T regime}
In order to understand the crumpled state at low tension, we introduce a specific form for the energy recently suggested to model regimes where both the thickness and the tension vanish~\cite{Chopin2015}:
\begin{equation}
U = U_{FT} + \frac{1}{2n}\gamma B \eta^{2 n},
\label{eq:Ucr}
\end{equation}
where $\gamma$ is a numerical prefactor and $n$ an adjustable exponent. The first term is the contribution of the stretching energy in the FT limit. The second term captures the contribution of the bending energy originating from a focusing of the curvature near the triangular tips. The value of $n$ can be inferred by inspection of the experimental data. For $n \neq 1$, $C/T$ would have a significant dependence with $\eta$ which is not observed, therefore we found that our data are best rationalized taking $n=1$ leading to : 
\begin{equation}
\frac{\overline{C}}{T} = \frac{1}{4} + \gamma \frac{B}{T}.
\label{eq:Ccr}
\end{equation}
 We can also note that the good agreement of the FT approach to model the contraction, even at large $\alpha$, indicates that the extra bending terms added in Eq.~\ref{eq:Ucr} does not depend on the tension leaving the scaling for $\eps$ unaffected. Taking $\gamma \approx 20$, data shown in Fig.~\ref{Fig7} are well adjusted by Eq.~\ref{eq:Ccr}.

\section{Conclusion}
In conclusion, we characterized experimentally and modeled the torque and contractile response of a soft Hookean ribbon submitted to large twist. In a stretched helicoid regime, a nonlinear torsional model based on the FvK equations quantitatively captures the nonlinear evolution of the longitudinal strain and torque with the twist. At large twist angle, above a wrinkling threshold, the ribbon exhibits wrinkles and folds responsible for a significant deviation from the helicoid response. We identified two regimes depending on the tension. In a high-$T$ regime, the macroscopic response is dominated by the external work of the load which can be quantitatively capture by a Far from Threshold approach which neglects the contribution of the folds. In a low-$T$ regime, bending contributions originating from the localized curvatures strongly affects the torsional response but not the contraction. This behavior at low tension can be captured by an extra bending term in the energy. This low-$T$ is interesting because the ribbon exhibits an exceptionally large contratability which may be useful for actuation in MEMS or in artificial muscles. More work is however needed to connect the global behavior with the local response to go beyond a phenomenological modeling.\\

\subsubsection{Acknowledgments}
We thank K{\'e}vin Flijane and Ga{\"e}tan Gourgaud for their help with the experiment and Vincent Demery, Arshad Kudrolli, and Mokhtar Adda-Bedia for stimulating discussions and critical readings of the manuscript. This work was funding by Ci{\^e}ncia Sem Fronteiras program (CNPq, Brasil) 313029/2013-05.

\end{document}